# Radiation-cooled Dew Water Condensers Studied by Computational Fluid Dynamic (CFD)

Clus Owen[1,2]; Ouazzani Jalil[3]; Muselli Marc[1,2]; Nikolayev Vadim[2,4,5], Sharan Girja[6], Beysens Daniel[2,4,5]

[1] Université de Corse, UMR CNRS 6134, Route des Sanguinaires 20000 Ajaccio, France
e–mail: clus@univ-corse.fr ; marc.muselli@univ-corse.fr
[2] OPUR International Organization for Dew Utilization, Paris, France
www.opur.u-bordeaux.fr
[3] Arcofluid, Bordeaux, France, e–mail: arcofluid@wanadoo.fr
[4] Commissariat à l'Energie Atomique, Grenoble
[5] Ecole Sup. de Physique et Chimie Industrielles, Paris, France
[6] Indian Institute of Management, Ahmedabad (India)

**Abstract.** Harvesting condensed atmospheric vapour as dew water can be an alternative or complementary potable water resource in specific arid or insular areas. Such radiation-cooled condensing devices use already existing flat surfaces (roofs) or innovative structures with more complex shapes to enhance the dew yield.

The Computational Fluid Dynamic – CFD – software PHOENICS has been programmed and applied to such radiation cooled condensers. For this purpose, the sky radiation is previously integrated and averaged for each structure. The radiative balance is then included in the CFD simulation tool to compare the efficiency of the different structures under various meteorological parameters, for complex or simple shapes and at various scales. It has been used to precise different structures before construction. (1) a 7.32 m² funnel shape was studied; a 30° tilted angle (60° cone half-angle) was computed to be the best compromise for funnel cooling. Compared to a 1 m² flat condenser, the cooling efficiency was expected to be improved by 40%. Seventeen months measurements in outdoor tests presented a 138 % increased dew yield as compared to the 1 m² flat condenser. (2) The simulation results for 5 various condenser shapes were also compared with experimental measurement on corresponding pilots systems: 0.16 m² flat planar condenser, 1 m² and 30° tilted planar condenser, 30 m² and 30° tilted planar condenser, 255 m² multi ridges, a preliminary construction of a large scale dew plant being implemented in the Kutch area (Gujarat, India).

**INTRODUCTION**

Dew condensation can be an interesting complementary renewable source of potable water for arid or insular areas (D. Beysens et al., 2003 and 2005; M. Muselli et al., 2002). Radiative cooled dew condensers are composed of a (white) low density Polyethylene plastic film including mineral fillers with high IR emissivity (T. Nilsson, 1996; produced by OPUR – www.opur.u-bordeaux.fr). The film is placed on styrofoam (polystyrene) for thermal insulation. This high radiative surface is passively cooled below the dew point temperature by radiative energy dissipation. Experimentations at the Ajaccio site (Corsica island, France) have been carried out for 8 years and various pilot systems have been tested. Their behaviour is now well understood and can be correlated with respect to a small number of meteorological parameters: wind speed $V$ (m s$^{-1}$), relative humidity $RH$ (%), cloud cover $N$ (octas), ambient temperature $T_a$ (°C), dew point $T_d$ (°C).

The description of the radiative condensers behaviour needs in particular the determination of the heat transfer coefficient surface/air. The heat transfer parameter can be calculated for planar surfaces with parallel air flow (V.S. Nikolayev, D. Beysens and M. Muselli, 2001; A.F.G. Jacobs, B.G. Heusinkveld and S. Berkowicz, 2004). However, the calculation of the heat exchange in complex outdoor radiative structures submitted to real wind is much more difficult. In addition, dew condensers works very often when the wind speed is quite small; what matters then is the tangential flow to the condenser, which is a mix up of free and forced convection. The relation between the wind speed as measured at 10 m above the ground and the air flow velocity tangential to a planar surface making an angle $\alpha$ with horizontal has been simulated by Beysens et al., 2003. The tangential velocity has been found minimum for $\alpha \approx 30°$, a result in agreement with the outdoor experimentation. However, experimental tests outdoor implies a large number of parameters, including the meteorological parameters. It needs a long time (usually one year) to average the season dependence and the results obtained with one geometry are difficult to extrapolate to another geometry. Then numerical experiments as carried on PHOENICS CFD Software can be very useful as a tool to determine the main characteristics of a new condensing structure, at least in a relative mode. These simulations permit to:

- Understand the thermal behaviour in limit conditions such as very weak wind speeds;

- Determine new condenser shapes. Numerical simulations can optimise new systems before building it outdoor;
- Predict the behaviour of new system when changing the scales (i.e. going from a mock up to a large system).

**PROGRAM SETUP**

A radiation-cooled condenser has to be simulated within three different aspects. (i) Thermal behaviour of the radiative material and the insulation material, including emissivity, conductivity and heat capacity. (ii) Radiative cooling power, a function of atmospheric conditions (sky emissivity, temperature, cloud cover) and condenser geometry. (iii) Incoming diffusive and convective (free or forced) heat from air flow, which depends on the wind speed and condenser geometry. The main contribution of CFD is to study the system by including all these interdependent parameters by means of an iterative calculation using the PHOENICS numerical code, based on the Finite Elements numerical method and the Navier–Stokes equations. It is noticeable that the calculation includes gravity and thus accounts for both free and forced convection.

**Radiative cooling**

The radiative power emitted from each cell depends on its local temperature. It is determined with a specific integration program performed on the following principle.

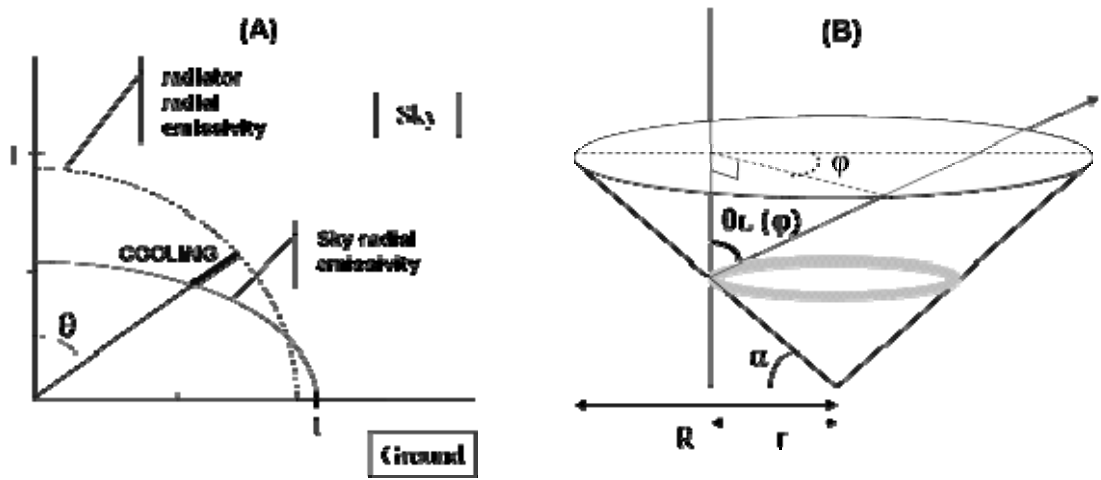

Fig. 1. (a) Variation of both radiator and sky emissivities with respect to the inclination angle $\theta$ (degree) with vertical. The bold line represents the neat cooling power (see text). (b) Integration performed for a funnel shape, from $\theta = 0$ to $\theta = \theta_L$ ($\theta$ limit), from $\varphi = 0$ to $\varphi = 360°$, from $r = 0$ to $r = R$. The result is weighted according to the surface of each funnel fraction (gray) with respect to the total funnel surface.

The relative cooling power is represented on Fig. 1a with respect to the angle $\theta$ (degree) inclination with vertical. It is the difference between the radiator emissivity (dashed line) and the sky angular dependant radial emissivity (full line). The sky angular emissivity is given by Berger et al., 2003:

$$\varepsilon_{s,\theta} = 1 - (1-\varepsilon_s)^{1/b\times\cos\theta} \tag{1}$$

$b = 1.66$

$\theta$ angle with zenith direction, $0 < \theta < \pi/2$

All calculations have been carried out for common night weather conditions in a temperate climate (France): clear sky, 288 K (15°C) ambient temperature and 80% relative humidity. The radiative balances of each elementary solid angle are then integrated as described in Fig .1b and for various tilted angle $\alpha$ (degree). The integrations are computed for various radiator temperatures. A $3^{rd}$ degree polynomial law is assumed to correlate the energy balance (W m$^{-2}$, related to the surface temperature) with $T_a$ (K) and $RH$ (%). In other words, each cell with temperature $T_c$ dissipates an energy $E_c$ that depends on $T_c$ and cell volume $V_c$.

**Computational Fluid Dynamic**

The PHOENICS software is suitable to run 3D or 2D simulations. The objects are placed in a framed space as presented on Fig. 2.

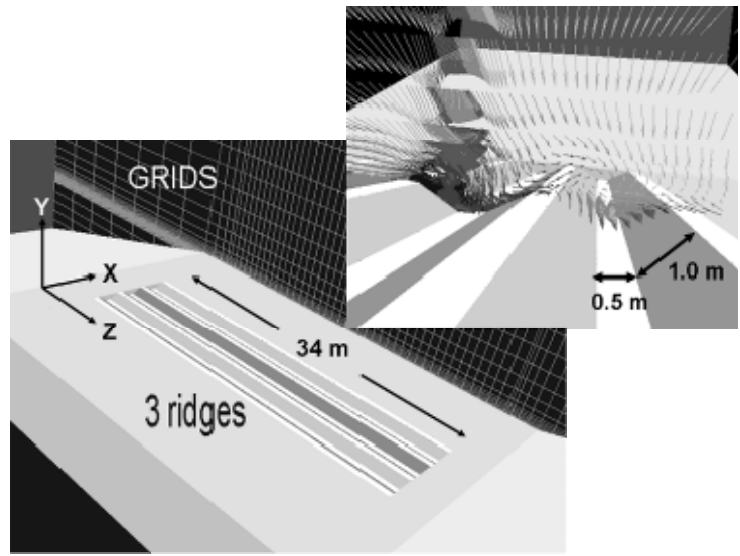

Fig. 2. Overview and detail of three ridges (255 m²) of the large scale condenser (India) as modeled in the virtual reality PHOENICS viewer.

The thermodynamic parameters are initialized for each cell center and cell sides. The radiative balance $E_c$ is added as input for each radiator elementary cell and a log type wind profile is given on one side of the simulated space that has been previously chosen as inlet. At each cell facing the inlet is given the wind velocity $V_c(y)$:

$$V_c(y) = V_{10} \ln(y/y_0) / \ln(10/y_0) \qquad (2)$$

Here $y_0$ (taken here to be 0.1 m) is the roughness length, $y$ (m) is the height of the cell center of the inlet and $V_{10}$ is the chosen 10m wind speed for each numerical experiment. The simulation gives the values of each interesting variable (Pressure $P$; Temperature $T_c$; fluid 3D velocity $V_x$, $V_y$, $V_z$), in a steady state and in function of the inlet velocity values. The efficiencies of condensers with different geometry and size can be then directly compared. For more accuracy in the comparison, all simulated shapes have the same radiators thickness (4 cm) and horizontal and vertical frames have the same size (2 cm).

**Data collection**

The software is a versatile tool as it allows the user access to any individual variable of any individual cell of the full space. Program sequences are inserted in suitable places in order to treat and sort the requested values. The condensation phenomenon has not been programmed. In this paper, a simple parameter will be discussed in order to compare the condensers efficiency: the mean surface temperature with a 15 °C ambient temperature. This value gives an immediate comparison of the structures efficiency. Only the upper cells of the radiator (in contact with the ambient air) are taken into consideration.

RESULTS

**Funnel shape simulation**

The (vertical) funnel shape reduces the free convection and then the heat exchange along the surface in blocking the heavier cool air at its basis, without any unfavourable wind direction because of its symmetrical behaviour. Cooling then should be increased and condensation enhanced. If we assess a symmetrical temperature distribution inside the funnel shape, any elementary surface is in radiative equilibrium with the facing condenser surface, so that the internal radiative budget is null. In addition, in masking the lower (and most IR emissive) atmospheric layer to most of the internal surface, the funnel shape lowers the intensity of downward long wave sky radiation and thus enhances the radiative cooling power.

Berger X. and Bathiebo J. (2003) estimated that the closest atmospheric layer from the ground contained in the first 15° solid angle is emitting 25° of the integral IR sky radiation. A. F.G. Jacobs, B.G. Heusinkveld and S.M. Berkowicz (2004) modelled and tested an inverted pyramid with the 4 condensation sides inclined at 30° from horizontal. A 3 cm thick styrofoam was used as insulation layer with the OPUR foil as the condensation surface. The area of the condensation surface was 1.11 m². The inverted-pyramid collector condensation gain as compared with a standard 1 m², 30° tilted planar condenser was measured close to 15 %.

A 7.32 m² funnel type condenser was simulated in this work. Lowering the cone angle reduces convection heating but also reduces radiative cooling. CFD is then a good tool to evaluate both effects and determine the best cone angle. Fig. 3 shows the funnel-shaped condenser equipped with an OPUR foil radiator together with its representation.

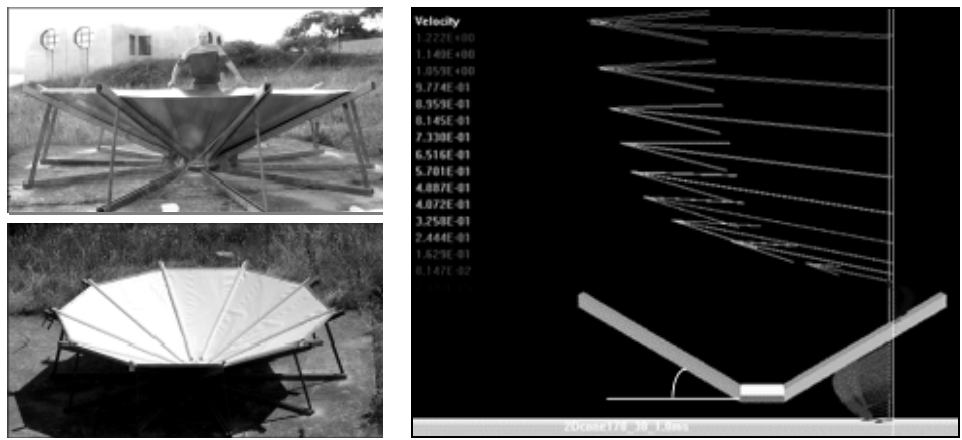

Fig. 3. Pictures of the funnel-shaped pilot (7.32 m²) and its corresponding 2D simulation. The internal surface is coated with OPUR Low Density PolyEthylene film insulated from below with 3 cm Styrofoam.

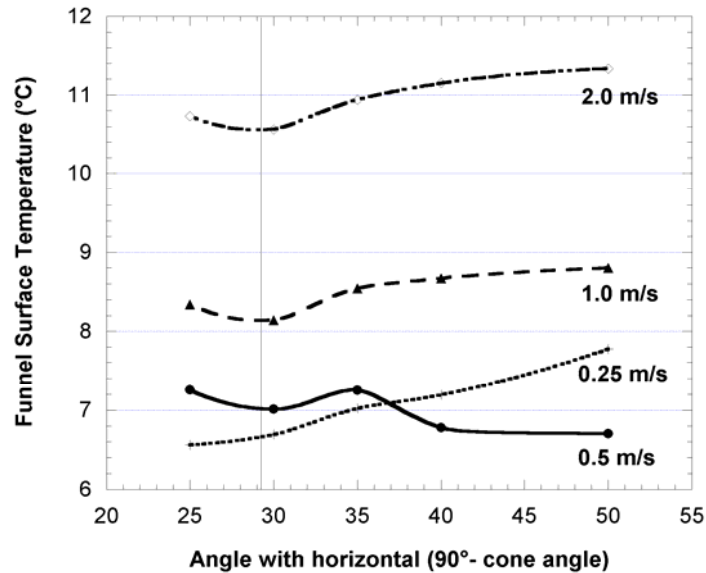

Fig. 4. Funnel surface temperature with respect to the angle with horizontal (90° - cone angle) for various wind speed (m s$^{-1}$, 10 m elevation). The 29° angle gives the best cooling efficiency (vertical line).

After the simulation of various angles (25°, 30°, 35°, 40° and 50°, see Fig. 4) a 29° angle with horizontal (60° cone angle) appears to give the best cooling efficiency. This is especially true for wind speeds > 1 m/s. For lower wind speeds, the air flow is in a mixed free/forced convection regime. The cooling efficiency can increase at large angles for moderate wind speed (0.5 m/s) although it decreases for lower speeds.

The choice of a 30° angle for the experimental funnel condenser was thus dictated by this study. Note that it is the same "best" angle as for plane condensers (D. Beysens et al., 2003). It also corresponds to an angle where the gravity forces that drives the condensed water flow for collection is decreased by only 50% with respect to the vertical case.

**Comparison between 5 dew condensers**

We now consider 4 experimental condensers (Fig. 5abcd) that have been studied in Ajaccio since 1998 and a large collector ("dew plant", Fig. 5e) in Panandro (NW India, see Sharan G. et al., 2006a; Sharan G., 2006b).

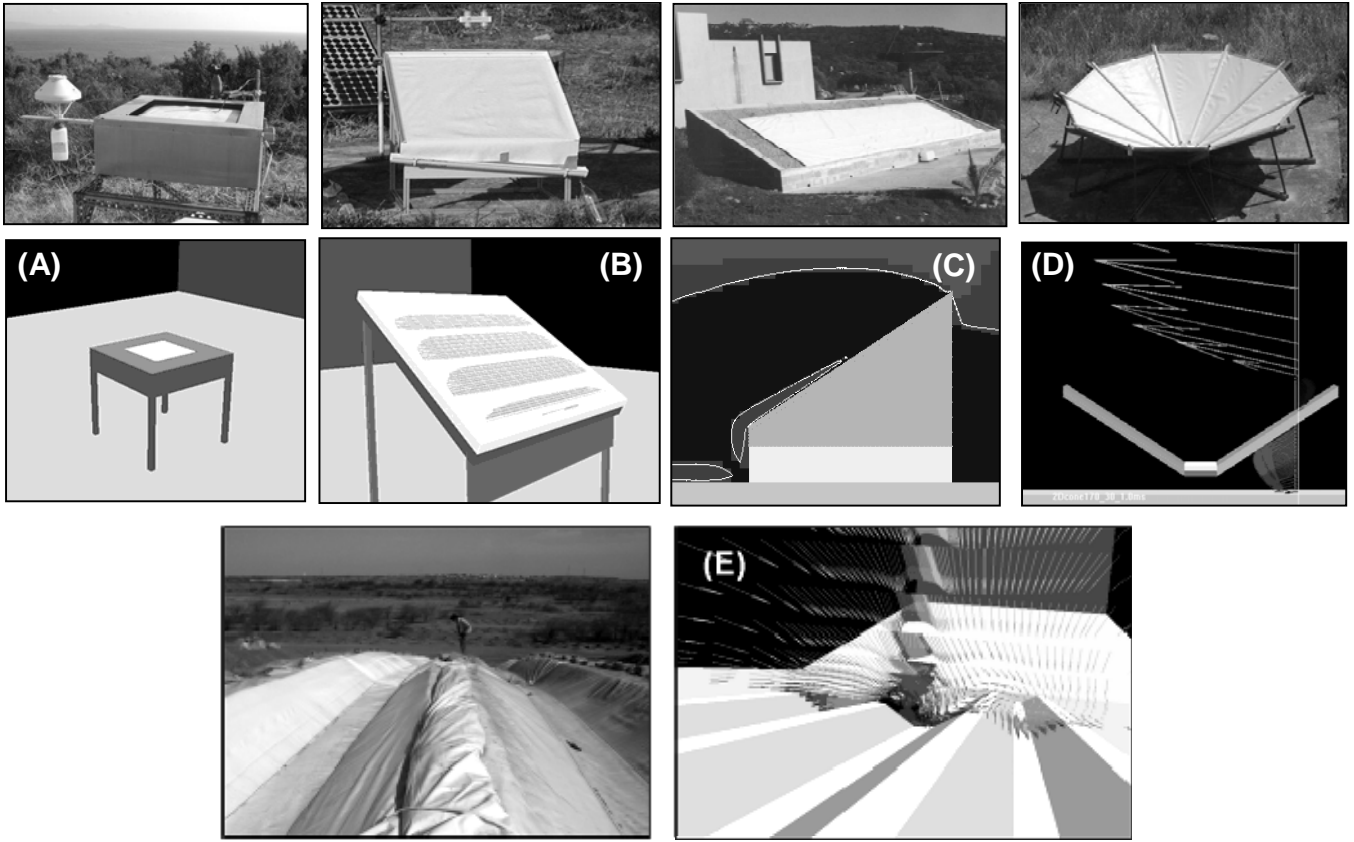

Fig. 5. Real (up) and virtual (down) condensers. (a) 0.4 x 0.4 m² isolated foil on a table. (b) 30° tilted with horizontal 1 x1 m² condenser, (c) 30° tilted with horizontal 3 x 10 m² condenser. (d) 7.32 m² funnel condenser with 60° cone angle. (e) Three trapezoidal ridges (top 50 cm, base 200 cm, two sides 30° tilted, height 50 cm, length 33 m on a 15° slope from horizontal), in Gujarat (NW India),

In the comparison of different condenser structures by simulation, it is anticipated that an increased condensation yield $h$ (mm/night) will correspond to an increased cooling efficiency. In Fig. 6 is shown the mean surface temperature $T_{cond} = <T_c>$ of the above structures simulated for the standard conditions as previously described. Surface temperature is given on the top face of each condenser's solid cell in contact with a fluid cell. (a, d) have symmetrical behavior in regard with wind direction. (b, c) have been orientated so as to expose their back to the dominant nocturnal wind. For the large scale ridge condenser, the simulation was performed with a horizontal meteorological wind coming from the top of the hill making a 30° angle with the slope axis, that is, 30° angle with the ridge axis. This orientation is close to the mean wind direction as measured during the dew events.

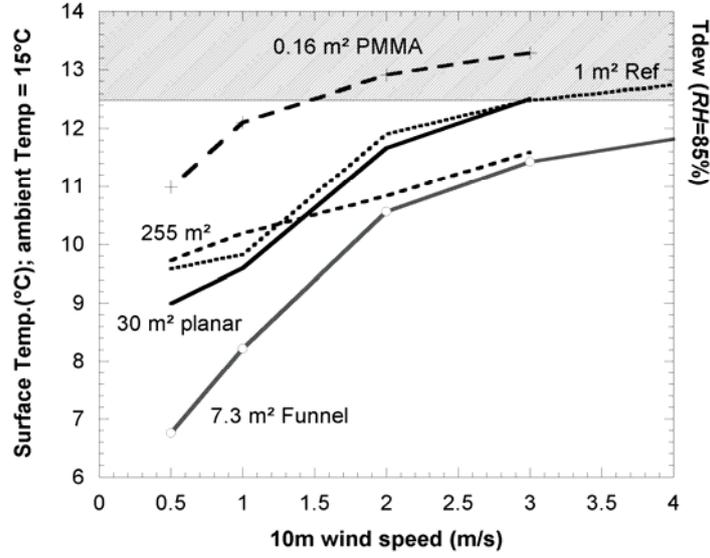

Fig. 6. Averaged surface temperatures obtained by numerical simulations and related to wind speed (10 m elevation). No condensation occurs in the gray area ($T_{cond} < T_d = 12.5$ °C, $RH < 85$ %).

The simulations are carried out without condensation ("dry air" conditions). However, it is possible to compare the cooling power of the 4 radiative cooled condensers with respect to a 1 m², 30° tilted, reference condenser. The relative efficiency of cooling factor or "temperature gain" $\Delta T_0$ can be defined (Beysens D. et al., 2003):

$$\Delta T_0 = \frac{T_{cond} - T_a}{T_{Ref} - T_a} \qquad (3)$$

Here $T_a = 15$°C and $T_{Ref}$ stands for the 1 m² condenser surface temperature.

The temperature gain (Fig. 7a) obtained from numerical simulations can reach 50%. The dew water yields depend mainly on the difference $T_{cond} - T_d$ or equivalently $RH$ (Lushiku, E.M. et al.,1989; Muselli M. et al., 2006) and the radiative cooling power is limited by the cloud cover $N$, all parameters that is not specific to the condenser geometry. The radiative budget and the air flow that are particular to each shape are included in the program. Then the comparison through the reduced $\Delta T_0$ factor of the different condensers for $N = 0$ (clear sky) and $RH = 80$ % can be extended to other meteorological situations.

These temperature simulation can thus be compared to experiments through a reduced dew yield $h_{cond}/h_{ref}$ ($h$ is in mm/night; see Fig. 7b). A good correlation is observed between the simulated $\Delta T_0$ and the experimental $h_{cond}/h_{ref}$ for 4 condensers (plane and funnel). The measurements on planar condensers were performed in Ajaccio from 09/07/2003 to 12/06/2003 (45 dew events) and the measurements on the cone, in Ajaccio from 05/25/2005 to 11/14/2006 (107 dew events) on the 1m² and cone condensers.

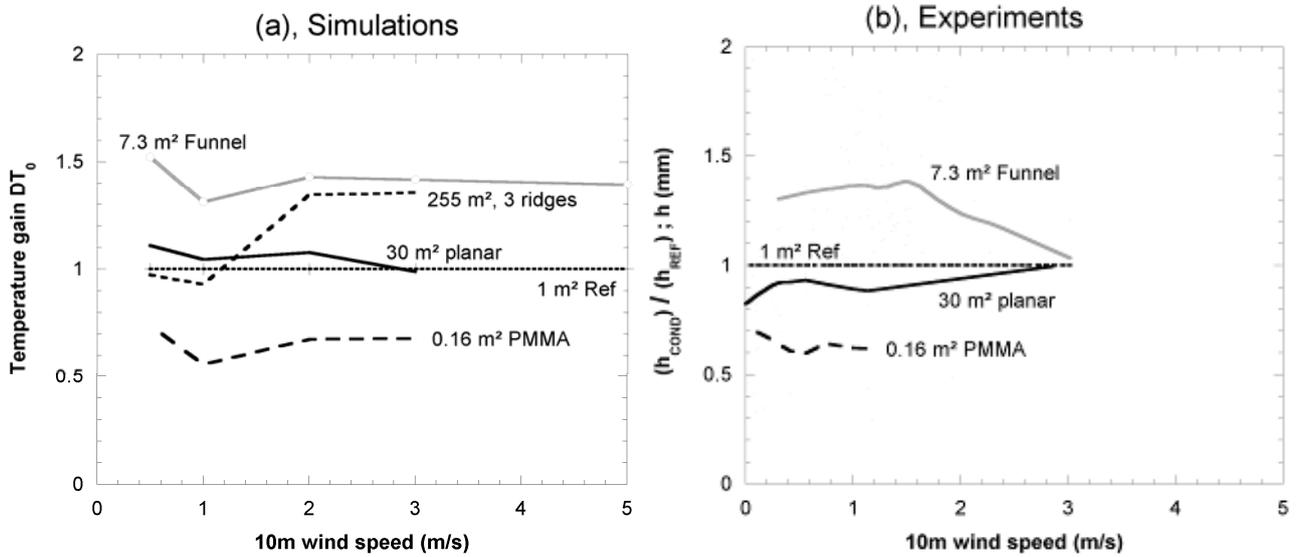

Fig. 7. (a), "temperature gain" or cooling factor $\Delta T_0$ obtained by numerical simulations for 5 various condensers systems from 0.16 to 255 m² and for assumed $Ta = 15°C$ and $RH = 80\%$. The 1 m², 30° planar condenser is taken as reference.

(b), "dew gain" or relative dew yields obtained in Ajaccio on 4 types of condensers and reported in function of wind speed (non available for the 255 m² condenser). $h$ is expressed in mm/night; data are smoothened by a 70% weighing function.

The noticeable correspondence between the calculated temperature gain and the experimental dew gain means that it is possible to reach a quantitative comparison of various condenser's shapes efficiency with this numerical tool. Condensation process is complex, involves several meteorological parameters presenting a high variability and models developed for estimation of the condensation yield are still incomplete. It is noticeable (Fig. 7b) that on experimental values, the 0.16 m² PMMA plate stops condensing for wind speeds above about 1.1 m/s, whereas the planar condensers stops above 3 m/s wind speed. The simulation of the horizontal plane condenser (Fig. 6) gives indeed less efficiency than the other structures, in agreement with the previous studies (D. Beysens et al., 2003). As a supplementary tool for comparison, Table 1 gives the yields of each condenser configuration as compared to the 1 m² 30° tilted planar condenser (cumulated during the whole measurement period). Note that the funnel shape gives a cumulated dew yield 38% larger the one with the 1 m² planar condenser along the same period.

Table 1. Yields of each condenser configuration compared to the 1 m² 30° tilted planar condenser (values cumulated during the whole measurement period).

| 1 m² 30° tilted planar | 0.16 mm² horizontal planar (PMMA) | 30 m², 30° tilted planar | 7.32 m², 60° angle cone funnel | 3 trapezoidal ridges, 255 m² |
| --- | --- | --- | --- | --- |

| | | | | | |
|---|---|---|---|---|---|
| <DT$_0$>, integration from 0 to 3 m/s | 1.00 | 0.65 | 1.05 | 1.40 | 1.15 |
| <Cumul dew X / cumul dew Ref> | 1.00 | 0.68 | 0.91 | 1.38 | 0.81 |

Some measurements were also performed in Panandro (Kutch area, Gujarat state, NW India) from 03/22/2006 to 04/20/2006 (22 dew events) on a 30° tilted 1 m$^2$ plane condenser and the ridge – type condenser. The comparison simulation - experiment is delicate as there are no wind measurements available for the ridges. In Fig. 8 is reported the dew gain as in Fig. 7b, however with respect to the 1 m$^2$ condenser dew yield). The mean wind speed previously measured from 09-01-2005 to 02-01-2006 at the same place was about 1.5 m/s at 10 m elevation during the dew events. This corresponds to the mean value 0.2 mm/night in Fig. 8. In both conditions (Fig. 7a, 1.5 m/s; Fig. 8, 0.2 mm/night) the ridge yield is about equal to the 1 m$^2$ condenser. Smallest $h$ values correspond to highest windspeed; here also the experiment (Fig. 8) and the simulation (Fig. 7a) show that the ridge condenser approaches the funnel shape efficiency. However, a clear difference between simulation and experiment is observed for $h > 0.2$ mm or windspeed > 1.5 m/s. As only three data are concerned and windspeed is lacking, no definite commitment can be made about this discrepancy. (One plausible hypothesis for highest condensation events is a superior fog involvement on aerial framed condenser exposed in the wind than on condensers built on the ground, even if they are larger. That unforeseen result is corroborated with numerous observations from 3 month passed on the field. An observation protocol has been adjusted in India in order to answer this interrogation.)

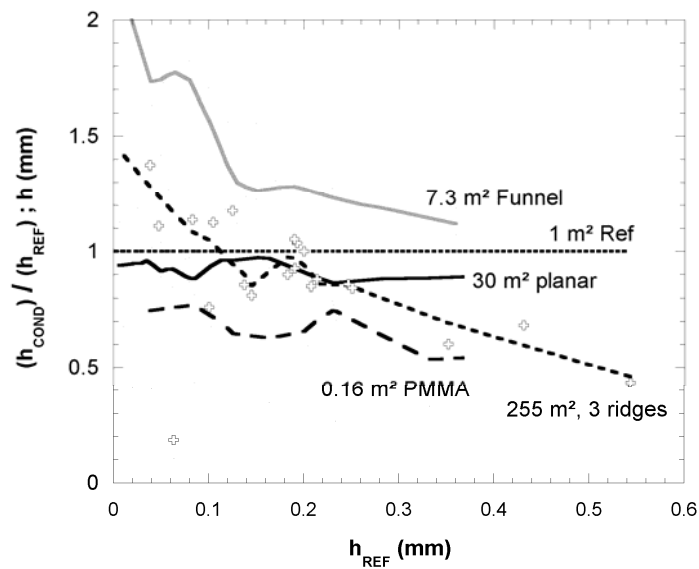

Fig. 8. Measurements of dew yields (mm/night) for all 5 condensers with respect to the planar 1 m² condenser. Data are smoothened by a 40% weighing function to enlight tendencies. Crosses represent the yields as measured on one ridge of the dew plant.

## CONCLUSION

A specific simulation program has been elaborated with the PHOENICS software to describe the functioning of radiative dew condensers. Five different geometries (0.16 m² planar horizontal, 1 m² and 30 m² planar at 30° with horizontal, 7.32 m² conical and a large scale 255 m² 3 ridges condenser) have been considered. The numerical simulations of cooling yields have been compared with the actual dew yields in the corresponding real dew condensers. The comparison is made possible by assuming a linear relationship between dew yields capability or "dew gain" and cooling below the ambient temperature efficiency (the simulation allowed an optimized orientation to be proposed for a 10 000 m² dew water plant that is under construction in NW India.)

## ACKNOWLEDGMENTS

This work has been started with the participation of Arcofluid and was supported partly by the National Agency for Innovation ANVAR-OSEO-Corse. The PhD of one of us (OC) is granted by the Collectivité Territoriale de Corse (CTC).

## REFERENCES

Berger X, Bathiebo J., 2003. Directional spectral emissivities of clear skies, Renewable Energy, 28(12):1925-1933.

Beysens D., Milimouk I., Nikolayev V., Muselli M., Marcillat J., 2003. Using radiative cooling to condense atmospheric vapour: A study to improve water yield, Journal of Hydrology 276:1-11.

Beysens D., Muselli M., Niklayev V., Narhe R., Milimouk I., 2005. Measurement and modeling of dew in Island, coastal and Alpine areas, Atmospheric Research, 73(1-2):1-22.

Beysens D, Ohayon C, Muselli M, Clus O., 2006. Chemical and bacterial characteristics of dew and rain water in an urban coastal area (Bordeaux, France). In press

International Organization for Dew Utilization – OPUR – www.opur.u-bordeaux.fr

Jacobs A.F.G., Heusinkveld B.G., Berkowicz S., 2004. Dew and Fog Collection in a Grassland Area, The Netherlands. Proceedings: Third International Conference on Fog, Fog Collection and Dew, Cape Town, South Africa.

Lushiku, E.M., Kivaisi, R.T., 1989. Optical properties of obliquely evaporated aluminium. Proceedings of SPIE 1149, 111–114.

Monteith, J.L., Unsworth, M.H., 1990. Principles of Environmental Physics, Second ed, Chapman & Hall, New York.

Muselli M., Beysens D., Marcillat J., Milimouk I., Nilsson T., Louche A., 2002. Dew water collector for potable water in Ajaccio (Corsica Island, France). Atmospheric Research, 64:297-312.

Muselli M., Beysens D., Milimouk I., 2006. A comparative study of two large radiative dew water condensers. J. of Arid Environment 64, 54-76

Nikolayev V.S., Beysens D., Muselli M., 2001. A computer model for assessing dew/frost surface deposition. Proceedings of the Second International Conference on Fog and Fog Collection, St John's (Canada) July 2001 Eds.R.S. Shemenauer and H. Puxbaum, IRDC, p.333 – 336

Nilsson T., 1996. Initial experiments on dew collection in Sweden and Tanzania, Solar Energy Materials and Solar Cells, 40:23-32.

Sharan G., Beysens D., Milimouk I., 2006a. A Study of dew Water Yields on Galvanized Iron roof in Kothara (North-West India). Journal of Arid Environment, in press

Sharan G., 2006b. Dew Harvest To Supplement Drinking Water Sources in Arid Coastal Belt of Kutch. Centre for Environmental Education, India. Book